\documentclass[runningheads]{llncs}

\usepackage[T1]{fontenc}

\usepackage{graphicx}

\usepackage{algorithm}

\usepackage[noend]{algpseudocode}

\begin{document}

\title{New Class of Ciphers Using Hardware Entropy Source}

\author{Jan J. Tatarkiewicz \and Wies{\l}aw B. Ku{\'z}micz}

\authorrunning{J.J. Tatarkiewicz and W.B. Ku{\'z}micz}

\institute{RANDAEMON sp. z o.o, ul. Ksawer{\'o}w 21, 02-656 Warsaw, Poland
\email{\{kuba.tatarkiewicz, wieslaw.kuzmicz\}@randaemon.com}\\
\url{https://www.randaemon.com}}

\maketitle

\begin{abstract}
We present a novel, computationally simple method of hiding any message in the stream of random bits by using a secret key. The method is called Bury Among Random Numbers (BARN). A stream of random bits is produced by extracting the entropy of a physical process in a hardware-based true random number generator (TRNG). The process of placing bits of a message into the stream of random bits is governed by the number of random bits skipped between subsequent insertions. The set of numbers that correspond to the steps of BARN is derived from a random number also provided by TRNG. Hence BARN cipher does not depend on any arithmetic function. We propose an effective method of computing random keys from a given number of random bits. We estimate the number of permutations that need to be tested during a brute-force attack on the new cipher for various key lengths. Some practical applications for the new class of symmetrical ciphers are discussed.

\keywords{cryptography \and symmetrical cipher \and hardware random numbers generator.}

\end{abstract}

\section{Introduction}
While testing a new type of random number generator, we accumulated a large set of random bits. A stream of bits was produced by extracting the entropy of a physical process in a hardware-based true random number generator (TRNG)~\cite{ref_article1}. Thinking about the ways these bits can be utilized, we noticed that it would be very easy to hide any message inside the set. Thus, steganographic cryptography\footnote{\emph{stegano} [Greek] covered, concealed; \emph{crypto} [Greek] hidden, secret.} was conceived. In this paper, we outline a very simple method of randomly inserting bits of a message into a continuous stream of random bits. The method is appropriately called Bury Among Random Numbers or BARN. In Section 2, the details of the encryption method are presented. Section 3 describes one of the possible methods of creation of keys by using a portion of a stream of random bits. Security analysis of the method is discussed in Section 4: cracking BARN resembles a search for a needle in the haystack. Section 5 summarizes the paper and outlines the possible applications of the method, especially when committed to silicon.

\section{Encryption algorithm}
Generally, any message $M$ in a digital world can be considered as a set of bits. Let $M$ have $\mu$ bits. Let $P$ be a stream of random bits generated by TRNG. The number of bits available from the stream is much larger than $\mu$. Assuming that a key $K$ consists of $\kappa$ natural, non-zero numbers, we can describe the BARN algorithm as a simple replacement of {\itshape i-th} bit $P_{i}$ in the stream by {\itshape j-th} bit $M_{j}$ of the message, where $j\in\{1, ..., \mu\}$, $t=j-\lfloor \frac{j-1}{\kappa}\rfloor*\kappa$, and:

\begin{equation}
 i = \lfloor \frac{j-1}{\kappa}\rfloor \times \sum_{l=1}^{l=\kappa} K_{l} + \sum_{l=1}^{l=t} K_{l}
\end{equation}
 
The function $\lfloor x \rfloor$ or {\itshape floor} is also called {\itshape integer} because it leaves an integer portion of the number $x$. If $\mu>\kappa$, then elements of a key $K$ are being reused cyclically. BARN algorithm is represented by a {\itshape pseudocode}:\\

\begin{algorithm}
\caption{BARN}\label{barn}
\begin{algorithmic}
\Require $P \gets \textit{stream of random bits}$
\Require $M \gets \textit{message, $\mu$ bits}$
\Require $K \gets \textit{encryption key, $\kappa$ elements}$
\State $k \gets 0$
\State $i \gets 0$
\State $j \gets 0$
\State \textit{loop}:
\State $j \gets j+1$
\If {$j > \mu$} \textbf{goto} \textit{end}
\State $k \gets \textit{k } \textit{mod } \textit{$\kappa$}+1$
\State $i \gets i+K(k)$
\State $P(i) \gets M(j)$
\EndIf
\State \textbf{goto} \textit{loop}
\State \textit{end}
\end{algorithmic}
\end{algorithm}

Visual representation of the insertion process shows how computationally simple the algorithm is. The BARN method starts with a stream of random bits:

\begin{figure}
\includegraphics[width=\textwidth]{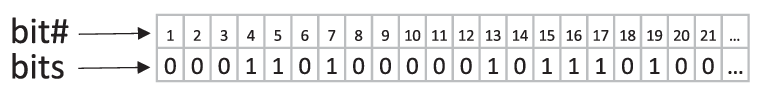}
\caption{Stream $P$ of random bits.} \label{fig1}
\end{figure}

For illustration purposes and to maintain clarity, we choose a rather short message $M$ of only 10 bits:

\begin{figure}
\includegraphics[width=\textwidth]{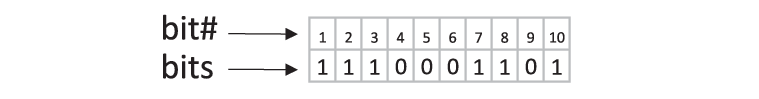}
\caption{Message $M$.} \label{fig2}
\end{figure}

In the next section, we will discuss methods of creating keys for BARN by using a stream of random bits and some elementary algorithms. Here, to simplify our example, we choose a very short key $K$ of just four, quaternary digits:

\begin{figure}
\includegraphics[width=\textwidth]{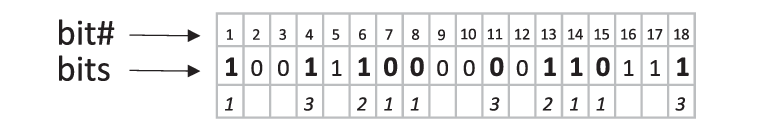}
\caption{Set of random bits $P$ with embedded message $M$ using key $K=\{1,3,2,1\}$.} \label{fig3}
\end{figure}

For $j=\mu$ Equation 1 provides the total length of the BARN cipher. In the example, $\kappa=4$, $\mu=10$ and $t=2$ so the length of the cipher is $18$. The numbers at the bottom of Figure~\ref{fig3} show how many bits are skipped between insertions - these are the elements of the key reused cyclically. We bolded inserted bits to emphasize the differences between a raw stream and the one with the message embedded. On average only half of the bits in the random stream are being changed by the insertion, e.g. the first bit was changed but not the fourth. This results from the basic property of a stream of random bits, i.e., zero and one are equally probable in such a stream. After insertion, the proportion of zeroes and ones in the cipher can change. In our example, among 18 random bits, there were 7 ones but after insertion, this number increased to 9 or (by chance) the expected, statistical average. When considering ASCII text files (7-bit encoding), the number of zeroes inserted will be slightly smaller than the number of ones. Basic ASCII characters including digits, brackets, etc., use 254 ones and only 229 zeroes. For UTF-16 (UNICODE~\cite{ref_article2}) files, this proportion will be different, i.e., there will be many more zeroes than ones because the encoding table contains scripts for many languages and many symbols, etc.

The knowledge of the key $K$ is sufficient to decode the BARN cipher - there is no need to have TRNG available. Thus, BARN is a symmetrical cipher like e.g. AES~\cite{ref_article3}. Unlike AES which is a block cipher, BARN is also capable of encoding continuous streams of information. However, BARN is not a typical stream cipher because it uses real TRNG (hardware source of entropy). Contrasting simple stream ciphers like e.g. GRAIN~\cite{ref_article4}, BARN is computationally even simpler: no arithmetic at all.

\section{Creation of keys}
The algorithm described in the previous section places few limitations on keys: a set of numbers $K_{i}$ forming the key should not contain only ones (such a key would copy the whole message into consecutive bits of a stream) and none of the elements $K_{i}$ can be zero (this would mean copying two bits of a message into the same bit of a stream). Typically, cf. Diffie-Hellman~\cite{ref_article5,ref_article6} public key system, the key is expressed in the number of bits that is a power of 2, e.g. 256-bit key. The keys should be as random as possible. We assume the availability of a stream of random bits from TRNG, so the method of creating a key starts with the number of random bits. For example, the set of numbers $K_{i}$ can be created from such a random string using different counting systems:

\begin{itemize}
\item ternary i.e., base 3 or digits $\{1,2\}$ will be used
\item quaternary i.e., base 4 or digits $\{1,2,3\}$ will be used
\item octal i.e., base 8 or digits $\{1,...,7\}$ will be used
\item decimal i.e., base 10 or digits $\{1,...,9\}$ will be used
\item hexadecimal i.e., base 16 or digits $\{1,...,15\}$ will be used
\end{itemize}

Each of the above-mentioned systems requires a different number of subsequent bits to denote a digit in the system:

\begin{itemize}
\item ternary: 2 bits, [01] and [10], efficiency 1/2
\item quaternary: 2 bits, [01], [10], and [11], efficiency 3/4
\item octal: 3 bits, [001], ..., [111], efficiency 7/8
\item decimal: 4 bits, [0001], ..., [1001], efficiency 9/16
\item hexadecimal: 4 bits, [0001], ..., [1111], efficiency 15/16
\end{itemize}

Here efficiency means the portion of random bits stream that can be used to create non-zero elements of a key. The number $\kappa$ of non-zero digits $K_{i}$ created on average from a given length of a random stream of bits is presented in Table~\ref{tab1}:

\begin{table}
\centering
\caption{Average number of elements for various keys.}\label{tab1}
\begin{tabular}{|l||c|c|c|c|c|c|}
\hline
 & {\itshape 64-bit} & {\itshape128-bit} & {\itshape 256-bit} & {\itshape 512-bit} & {\itshape 1024-bit}\\
\hline
{\itshape Ternary} & 16 & 32 & 64 & 128 & 256\\
{\itshape Quaternary} & 24 & 48 & 96 & 192 & 384\\
{\itshape Octal} & 18 & 37 & 74 & 149 & 298\\
{\itshape Decimal} & 9 & 18 & 36 & 72 & 144\\
{\itshape Hexadecimal} & 15 & 30 & 60 & 120 & 240\\
\hline
\end{tabular}
\end{table}

Our algorithms for creating keys $K$ from random bits are simple computationally. However, each counting system leads to a quite different length of a cipher depending on the average value of the elements of a key:

\begin{itemize}
\item ternary: average $M$ length enlargement by a factor of 1.5
\item quaternary: average $M$ length enlargement by a factor of 2
\item octal: average $M$ length enlargement by a factor of 4
\item decimal: average $M$ length enlargement by a factor of 5
\item hexadecimal: average $M$ length enlargement by a factor of 8
\end{itemize}

For example, the message $M$ with the length of $\mu$ bits will on average require $4\mu$ random bits with an octal key but only $2\mu$ random bits for a quaternary key. For the same length of a string of random numbers used to create these two keys, the octal one will provide better security, because of a larger number of possible permutations. Combining the above-mentioned enlargements with the number of permutations given in Table~\ref{tab2}, it is clear that different schemes will be suitable for various applications of the BARN:

\begin{table}
\centering
\caption{Average number of permutations for various types and lengths of keys.}\label{tab2}
\begin{tabular}{|l||c|c|c|c|c|c|}
\hline
 & {\itshape 64-bit} & {\itshape128-bit} & {\itshape 256-bit} & {\itshape 512-bit} & {\itshape 1024-bit}\\
\hline
{\itshape Ternary} & 6.55E+004 & 4.29E+009 & 1.84E+019 & 3.40E+038 & 1.16E+077\\
{\itshape Quaternary} & 2.82E+011 & 7.98E+022 & 6.36E+045 & 4.05E+091 & 1.64E+183\\
{\itshape Octal} & 1.63E+015 & 1.86E+031 & 3.45E+062 & 8.31E+125 & 6.91E+251\\
{\itshape Decimal} & 3.87E+008 & 1.50E+017 & 2.25E+034 & 5.08E+068 & 2.58E+137\\
{\itshape Hexadecimal} & 4.38E+017 & 1.92E+035 & 3.68E+070 & 1.35E+141 & 1.83E+282\\
\hline
\end{tabular}
\end{table}

\section{Security analysis}
John von Neumann famously wrote~\cite{ref_article7}: \textit{Any one who considers arithmetical methods of producing random digits is, of course, in a state of sin. For, as has been pointed out several times, there is no such thing as a random number - there are only methods to produce random numbers, and a strict arithmetic procedure of course is not such a method.} This statement excludes arithmetical procedures from generating good random numbers but it also gives a clear advantage to hardware TRNG-based cryptography. There is no method of breaking the cipher other than a brute-force guessing of all possible keys. One example of such a guess for the cipher in Figure~\ref{fig3} is presented in Figure~\ref{fig4}:

\begin{figure}
\includegraphics[width=\textwidth]{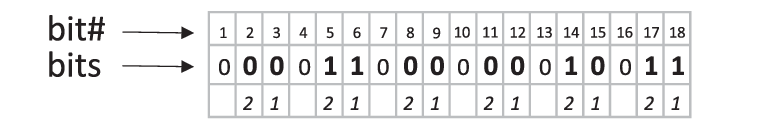}
\caption{Decryption with guessed key $K=\{2,1,2,1\}$.}\label{fig4}
\end{figure}

The message deciphered with a guessed key $K=\{2,1,2,1\}$ happens to be [001100001011] and is not similar to the original message [1110001101] at all, cf. Figure~\ref{fig2} where the original key was $\{1,3,2,1\}$. We assumed that an attacker knows everything about the system, even the type and random bit length of the original key but not the key.

A large number of possible keys in the BARN creates a large number of possible messages that can be guessed. Without knowing the original message, it is hard to decide which is the real one unless the attacker can toss the same known message multiple times into the system. Even if this is possible, one can only guess the key with some level of probability, depending on the number of times the very same text is being encrypted. If the transmitted messages always start with the same sequence of bits, as it would be with a standard RTF or PDF file, there is a non-zero probability of guessing the initial elements of the reusable key but not the whole one. This holds for the keys that are longer than the fixed header of the files.

The simplicity of BARN allows for security measures that can be added to practical implementations, especially for low-power devices like Internet-of-Things (IoT). Practical testing of BARN showed that cipher sequences analyzed statistically appear as not-so-good random sequences without any indication of which bits were replaced. We did extensive testing of various plaintext inputs and we found that BARN-encrypted messages are statistically indistinguishable from random numbers simply because statistical tests reveal bad randomness only for very long sequences of bits as a whole and not in the smaller sections. Statistical analysis of various sequences generated with a wide range of key types and lengths shows only slightly worse randomness of the cipher vs. real random sequence and this does not allow for any decryption unless brute force checking of all possible permutations of keys is tested.

\subsection{Probabilities}
Traditionally the number of permutations that need to be checked during a brute-force attack on the cipher is expressed as the power of 2, hence Table~\ref{tab2} can be rewritten with the powers of 2 that represent the average number of permutations for a given key:

\begin{table}
\centering
\caption{The powers of 2 that approximate the number of permutations in Tab.~\ref{tab2}.}\label{tab3}
\begin{tabular}{|l||c|c|c|c|c|c|}
\hline
 & {\itshape 64-bit} & {\itshape128-bit} & {\itshape 256-bit} & {\itshape 512-bit} & {\itshape 1024-bit}\\
\hline
{\itshape Ternary} & 16 & 32 & 64 & 128 & 256\\
{\itshape Quaternary} & 38 & 76 & 152 & 304 & 608\\
{\itshape Octal} & 50 & 103 & 207 & 418 & 836\\
{\itshape Decimal} & 28 & 57 & 114 & 128 & 456\\
{\itshape Hexadecimal} & 58 & 117 & 234 & 468 & 937\\
\hline
\end{tabular}
\end{table}

The numbers listed in Table~\ref{tab3} are very large indeed even for the short lengths of the keys. For example, if the attacker can process $10^9$ keys per second (multi GHz processor), a 256-bit ternary key would require more than $584$ years to run through all possible messages - this time does not include checking which messages make sense. We would like to point out again that lengthening the key does not increase the computing load on the processor that encodes or decodes the BARN cipher. Hence choosing even a longer key in the above example for the ternary key requires the same computing power with just a little bit more memory to store the longer key.

\section{Summary}
We showed in previous sections that BARN could be a very effective cipher providing that real TRNG is available. BARN method and creation of possible keys from a series of random numbers allow embedding into IoT devices with limited computing resources. If TRNG can be placed on a chip, then the whole system can be part of the System-on-Chip, which is the most secure solution. There are various instances when different TRNGs are needed, i.e., TRNGs that can generate different numbers of random bits per second. For the simplest solutions, as previously mentioned, e.g. encoding of short text messages, a low throughput TRNG of about 12 kbps would be enough. Cell phones limit sound bandwidth to about 8 kbps by using G.729 codec~\cite{ref_article8}. BARN ternary keys would enlarge encoded sound files by a factor of 1.5. This means that BARN cipher with ternary keys and using 12 kbps TRNG will also suffice there. On the other hand, for the encoding of files on the cloud, one would need TRNG with a much higher throughput. For example, to encode 500 Mbps with the quaternary keys, one would need TRNG that could generate random numbers at speeds of the order of 1 Gbps. Such fast TRNGs would suffice to encode multiple video streams for Video-On-Demand of movies to individual consumers. UltraHD video with a resolution of 3840x2160 pixels requires a bandwidth of about 25 Mbps. 1 Gbps TRNG can service at least 20 such streams simultaneously when ternary keys are used. Practically, TRNGs with a total output of several Gbps can be mounted on a specialized card, directly supplying enough random bits to a single server. In between the above-described two extreme situations, TRNGs with moderate throughputs of random bits and using octal keys can be deployed for better security required by government, military, medical, etc. services.

We believe that the BARN method outlined in this paper can fill the niche that presently is missing some practical cryptography methods because of high computing needs for proven, secure ciphers like AES~\cite{ref_article3}, especially in the post-quantum cryptography, see e.g. projects sponsored by NIST~\cite{ref_article9}.

\subsubsection{\ackname}
We are very grateful to dr G. Ja{\'n}czyk, dr D. Szmigiel and their teams from IMiF (Warsaw, Poland), and dr K. Siwiec and dr J. Jasi{\'n}ski from ChipCraft (Lublin, Poland) for their expert work on our “proof of concept” true random number generators.

\end{document}